\documentclass[aps,prl,twocolumn,showpacs,superscriptaddress]{revtex4}

\usepackage{amsmath}
\usepackage{amssymb}
\usepackage{graphicx}
\usepackage{bm}

\begin{document}

\title{Interaction-induced localization of anomalously-diffracting nonlinear waves}

\author{Y. Linzon}
\affiliation{School of Physics and Astronomy, Faculty of Exact
Sciences, Tel-Aviv University, Tel Aviv 69978, Israel}

\author{Y. Sivan}
\affiliation{School of Physics and Astronomy, Faculty of Exact
Sciences, Tel-Aviv University, Tel Aviv 69978, Israel}

\author{B. Malomed}
\affiliation{Department of interdisciplinary studies, Faculty of
Engineering, Tel-Aviv University, Tel Aviv 69978, Israel}

\author{M. Zaezjev}
\affiliation{Universite' du Quebec, Institute National de la
Recherche Scientifique, Varennes, Quebec J3X 1S2, Canada}

\author{R. Morandotti}
\affiliation{Universite' du Quebec, Institute National de la
Recherche Scientifique, Varennes, Quebec J3X 1S2, Canada}

\author{S. Bar-Ad}
\affiliation{School of Physics and Astronomy, Faculty of Exact
Sciences, Tel-Aviv University, Tel Aviv 69978, Israel}

\date{\today}

\begin{abstract}

We study experimentally the interactions between normal solitons
and tilted beams in glass waveguide arrays. We find that as a
tilted beam, traversing away from a normally propagating soliton,
coincides with the self-defocusing regime of the array, it can be
refocused and routed back into any of the intermediate sites due
to the interaction, as a function of the initial phase difference.
Numerically, distinct parameter regimes exhibiting this behavior
of the interaction are identified.

\end{abstract}


\maketitle

Solitary waves in general, and optical solitons in particular, are
the most important manifestations of the interplay between
nonlinear self-focusing and the diffraction/dispersion, which are
inescapable characteristics of linear wave-packet propagation.
Periodic structures, such as weakly-coupled waveguide arrays
(WGAs) introduce two new important features to this problem.
First, they support discrete solitons (DS), which are robust
against collapse with increasing input power and propagation
length \cite{Christodoulides2,Eisenberg}. Second, they allow
control over the diffraction properties of tilted beams, giving
rise to regions of normal, zero or \emph{anomalous} diffraction
for linear excitations \cite{Roberto1}, and accordingly, regions
of self-focusing (SF) or self-defocusing (SDF) for nonlinear
excitations \cite{Roberto2}. These controllable characteristics
not only make such structures favorable candidates for
applications in optical switching. They also make optical WGAs a
model for a diverse array of classical and quantum systems that
have nontrivial dispersion properties and support localized
nonlinear modes, from Bose-Einstein condensates in periodic
optical lattices \cite{Kivshar,Boris}, to localized lattice
vibrations \cite{Flach&Willis} and nonlinear excitations in arrays
of mesoscopic-mechanical cantilevers \cite{Sievers}.

Recently, there is a growing interest in interactions between
co-propagating beams in these WGAs \cite{Christodoulides1,Aceves},
including interactions between spatially-separate solitons
\cite{Stegeman&Segev,Meier1,Greece}, as well as between
co-propagating solitons and linear waves
\cite{Meier2,Meier3,Todd,Flach,Russel}. In homogeneous Kerr media,
mutually coherent in-phase and out-of-phase solitons attract or
repel each other, respectively \cite{Stegeman&Segev}. It was found
theoretically \cite{Christodoulides1} and experimentally
\cite{Meier2,Meier3} that the collision between a DS and a weak
non-diffracting signal beam can lead to dragging of the blocker
soliton towards the weak signal beam. In Ref. [12], Meier
\emph{et. al.} have experimentally studied the interaction between
two narrow collinear DS in Kerr-nonlinear WGAs fabricated of
AlGaAs. It was found that a pair of DSs, separated by 3 sites, are
stable (\emph{i.e.} noninteracting) when the relative phase
difference between them is around $\pi$, and unstable (\emph{i.e.}
attractive) when it is close to zero, in accordance with
theoretical predictions \cite{Christodoulides1,Aceves,Todd}.
However, interactions involving {\em tilted} beams, at different
diffraction regimes, have never been studied experimentally.

\begin{figure}[t]
\includegraphics[width=8cm]{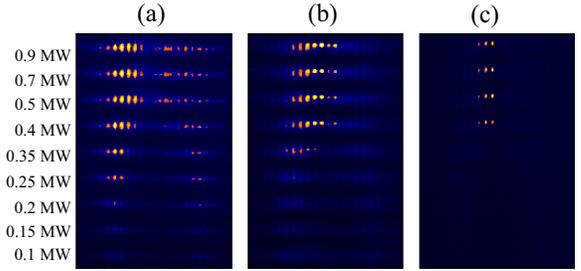} \caption{(Color online) Formation of a single normally injected
soliton in the glass waveguide array, for a 40 $\mu $m wide input
beam. The three panels correspond to different imaging conditions,
and in each one the input power increases from bottom to top. (a)
shows  images of the output facet of the sample with the imaging
lens adjusted for the laser wavelength. At low powers the
discrete-diffraction pattern is observed; At high powers a soliton
forms, but is out of focus. (b) The soliton becomes visible at a
different focal plane, corresponding to the Raman-shifted
wavelength, but its image is saturated. (c) The same image as in
(b), with an ND2 filter (X100 attenuation), showing that the
soliton is much stronger than the background. Setting (c) is used
to study soliton interactions.}
\end{figure}

In this paper, we report results for the interaction between a
normally-injected soliton (NS) and a nonlinear tilted beam (TB)
that propagates \emph{away} from the NS. We find that the
characteristics of the interaction are considerably different in
the SF and SDF regimes of TB propagation. Specifically, in the SDF
regime we demonstrate for the first time that it is possible to
\emph{restore} self-focusing to the TB, and to redirect it into
any of the intermediate sites, by varying the relative phase
difference.

We use a WGA fabricated of silica glass, as in Ref. [19], with the
periodicity $d=12$ $\mu $m and coupling constant $C=0.23$
mm$^{-1}$. The sample is 13 mm long, corresponding to 1.8 coupling
lengths. As the Kerr coefficient in glass is $\sim$500 times
smaller than in AlGaAs \cite{Stegeman&Segev}, shorter pulses with
higher peak powers are required to generate nonlinear effects.
Importantly, the lowest-order nonlinear loss mechanism in glass is
six-photon absorption, which has an extremely low probability
amplitude. DSs in glass WGAs, including TBs, are therefore
sustained at high powers without breakup \cite{Dima}, allowing a
broader range of powers and tilts to be explored. Glass is also
characterized by strong stimulated Raman scattering, which
expresses itself as self frequency-shift of the soliton's spectrum
to longer wavelengths \cite{Russel}. The formation of a soliton is
therefore accompanied by a change of the focusing plane, due to
the chromatic aberration of the output microscope objective lens.
This sudden change of focus enables us to distinguish between
strong solitary features and the linear background without
explicit spectral measurements, as demonstrated in Fig. 1.

\begin{figure}[t]
\includegraphics[width=8cm]{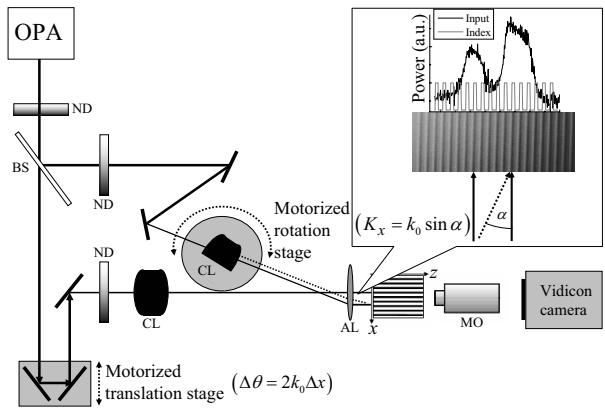} \caption{Experimental setup. OPA - optical parametric amplifier,
BS - beam splitter, ND - neutral density filter, CL - cylindrical
lens, AL - aspherical lens, MO - microscope objective. Inset:
Lateral intensity profile of the input beams. The normal beam
excites $\simeq $3-4 sites, while the tilted beam excites $\simeq
$4-5 sites. The excitation parameters, $\Delta\theta$ and $K_{x}$,
described in the text, are determined from applied translations
and rotations, respectively, using the free-space wavenumber
$k_{0}=2\protect\pi$/$\lambda_{0}$.}
\end{figure}

The experimental setup for the 2-beam experiment is sketched in
Fig. 2. A train of 70-fs pulses at $\lambda_{0}$=$1.52$ $\mu $m,
with $\leq$ 20 MW peak power, was split into two identical beams.
The normally injected beam forming the NS was given an adjustable
phase delay $\Delta\theta$ using a variable delay stage, and
shaped using a cylindrical lens. The second beam passed through a
rotating cylindrical lens that moved the beam on the aspherical
coupling lens, and thus changed its angle of incidence $\alpha$ at
the input facet of the sample. This angle determines the
transverse wavenumber $K_{x}$ and thus the phase difference
between adjacent sites in the TB excitation. Both beams were
coupled into the sample through the same aspherical lens, which
has a large numerical aperture ($0.68$ NA) and is optimized for
the operating wavelength. A unique feature of this setup is the
large clear aperture of the coupling lens, which in combination
with its small focal length, results in a wide range of accessible
incidence angles for the tilted beam, while maintaining
paraxiality conditions and good coupling efficiency. The edge of
the first Brillouin zone for this sample is attained at an
incidence angle of $\alpha =\arcsin (\lambda
_{0}/2d)=3.6^{\mathrm{o}}$. In order to have sufficient resolution
in $K_{x}$, that allows probing of the different diffraction
regimes for the TB, the lateral width of the input TB had to be
larger than $\sim $30 $\mu $m. The combined input intensity
profile for the two beams is shown in the inset of Fig. 2. The TB
input profile did not vary under relevant tilts.

\begin{figure}[t]
\includegraphics[width=8cm]{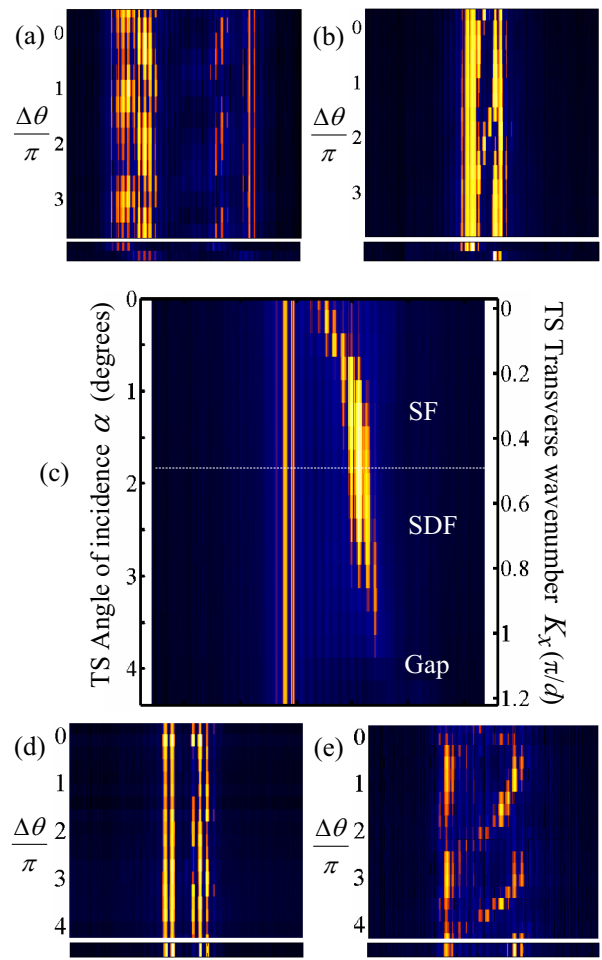} \caption{(Color online) (a),(b) Output profiles as a function of
the phase difference, with both beams at normal incidence. The
input powers are (a) $0.15$ MW, and (b) $0.7$ MW (Linear and
soliton regimes, respectively). (c) The output profiles of
$\emph{noninteracting}$ 0.7 MW beams as a function of the TB tilt
(data obtained with a very large time delay between the pulses).
(d),(e) Interactions of 0.7 MW beams as a function of the phase
difference for the tilt angles of (d) $0.6^{\mathrm{o}}$
($K_{x}=0.17\protect\pi $/$d$), and (e) $3^{\mathrm{o}}$
($K_{x}=0.84\protect\pi $/$d$). The bottom parts in (a),(b),(d)
and (e) show the output in the absence of interaction.}
\end{figure}

\begin{figure*}[t]
\includegraphics[width=11.5cm]{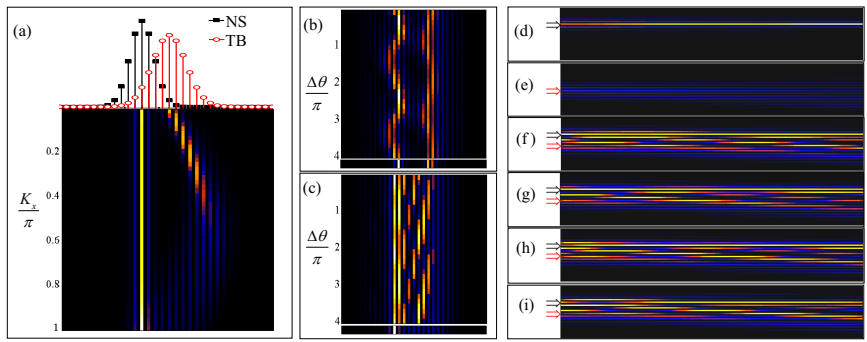} \caption{(Color online) Results of numerical simulations. (a) Top:
the amplitude distribution in the input, $|\Phi _{n}(0)|$. Bottom:
output intensity distributions as a function of the tilt $K_{x}$
for noninteracting beams (\emph{i.e.}, superposition of individual
solutions). (b),(c) Output intensity distribution as a function of
the phase difference for: (b) $K_{x}=0.24\protect\pi $ and (c)
$K_{x}=0.9\protect\pi $. (d)-(i) Propagation maps of the intensity
distribution as a function of $z$ for $0<z<13$ mm and
$K_{x}=0.9\protect\pi $: (d) NS without TB; (e) TB without NS;
(f)-(i) both beams present, with the relative initial phase
differences: (f) $0$, (g) $\protect\pi /2$, (h) $\protect\pi$, and
(i) $3\protect\pi /2$.}
\end{figure*}

Experimental results are shown in Fig. 3. Panels (a),(b) show the
results obtained when the two beams are at normal incidence. At
low input powers (a), linear interference fringes are superimposed
on the superposition of the two overlapping discrete-diffraction
patterns. It was found that the slope of the fringes exactly
coincided with the phase slope imposed by the translation motor.
At high input powers (b), the anticipated nonlinear instability
around $\Delta\theta =2\pi$, which partially diverts the beams
into intermediate sites, is observed \cite{Meier1}.

As tilts are imposed on the right DS (Fig. 3(c)), it is routed
aside. As expected, the sideways shift of the beam increases in
the SF regime, and decreases again in the SDF regime, where the
beam becomes delocalized (The data in Fig. 3(c) was obtained with
a very large time delay between the pulses, which inhibits the
interaction). Panels (d) and (e) of Fig. 3 show examples of
interacting NS and TB in the SF and SDF regimes of the TB,
respectively. For small tilts that correspond to a TB in the SF
regime (Fig. 3(d)), the observed attractive instability near
$2\pi$ phase multiples is less pronounced compared to the
collinear excitation (Fig. 3(b)). This may be expected due to the
decreasing overlap of the two fields as the TB is routed aside.
However, when the tilt angle is in the SDF regime of the TB (Fig.
3(e)), the instability is extended for phase differences far
beyond a vicinity of $2\pi$. Remarkably, the TB can be routed as a
localized DS into any of the intermediate sites, with the
appropriate phase setting, while the NS is partially annihilated.
Evidently, the WGA favors interaction when one of the beams
propagates in the SF regime and the other one in the SDF regime,
as demonstrated by the extended range of phases exhibiting
interaction dynamics.

In order to gain better understanding of the interaction dynamics,
we solved the one-dimensional discrete nonlinear Schr\"{o}dinger
equation in dimensionless form \cite{Christodoulides2,Musslimani},

\begin{equation}
i\frac{d\Phi _{n}}{dz}+h^{-2}(\Phi _{n+1}+\Phi
_{n-1}-2\Phi_{n})+|\Phi_{n}|^{2}\Phi _{n}=0,
\end{equation}

\noindent on a grid of 101 sites; $h^{-2}=Z_{NL}C$, the ratio
between the characteristic nonlinear length $Z_{NL}$ and the
diffraction length $Z_{D}=1/C$, is a single parameter
characterizing the sample. The input is taken as two Gaussians
centered at waveguide numbers $n_{c1}=51$ and $n_{c2}=55$,\\*
\begin{align}
\Phi_{n}(0)=\frac{A_{\mathrm{NS}}}{\sqrt{P^{\ast}}}\exp\left(-\frac{(n-n_{c1})^{2}}{(w_{1})^{2}}\right) \label{Eq2} \nonumber \\
+\frac{A_{\mathrm{TB}}}{\sqrt{P^{\ast}}}\exp\left(-\frac{(n-n_{c2})^{2}}{(w_{2})^{2}}+iK_{x}n+i\Delta\theta\right),
\end{align}
\noindent where $P^{\ast}$ is the characteristic soliton power
that defines $Z_{NL}$ \cite{Musslimani}. We took $P^{\ast}=0.5$
MW. This value and the other experimental parameters (\emph{e.g.}
the diffraction length, mode area, and Kerr coefficient) yield
$h^{-2}=4.9\times 10^{-3}$. The input widths, $w_{1}=2.5$ and
$w_{2}=3.5$, correspond to the experimental conditions. Finally,
in these units the first band of linear waves corresponds to
$0<K_{x}<\pi$.

We have found that, to reproduce the propagation dynamics observed
in the experiment, two conditions must hold regarding the
excitation amplitudes. First, the TB amplitude, for the above
input width, must be $45<A_{\mathrm{TB}}<60$. For higher
amplitudes (with $A_{\mathrm{NS}}=0$) the steering of the TB to
nearby sites in the SF region is suppressed, and the soliton
instead becomes trapped in one waveguide
\cite{Aceves,Bang&Miller}. Simultaneously, it becomes unstable
under small variations of the tilt. The second condition pertains
to the NS. While the excitation amplitude must be
$40<A_{\mathrm{NS}}<90$ for a NS to form by itself (\emph{i.e.}
with $A_{\mathrm{TB}}=0$), we have found that it must be
comparable to the TB amplitude when the two excitations are sent
together with the initial condition (2). Otherwise, the TB
vanishes during the propagation, with only a small effect on the
NS. In addition to the above constraints on the excitation
amplitudes, we have also found that the balance between the
strength of diffraction and nonlinearity must be in favor of the
nonlinearity, in the form of a small $h^{-2}$ parameter, to avoid
breakup of the TB into filaments due to modulational instability
\cite{Meier4}. Indeed, all of the above conditions were fulfilled
in our experiment.

Results of the numerical simulations are shown in Fig. 4, for
$A_{\mathrm{NS}}=48$ and $A_{\mathrm{TB}}=40$. Figure 4(a) shows
the amplitude profile of the input beam and the output intensity
distribution for individual (non-interacting) beams, as a function
of the TB tilt. Figures 4(b) and 4(c) show the output intensity
distribution as a function of the initial phase difference, for a
TB that is SF and SDF, respectively. The calculated results are in
good qualitative agreement with the experimental data displayed in
Fig. 3. In particular, Fig. 4(c) indeed shows routing of a
relocalized TB to nearby sites, which changes gradually over the
entire range of the phase difference, $0<\Delta\theta <2\pi $. The
propagation maps as a function of $z$, panels (d)-(i), demonstrate
how the NS is re-phasing the otherwise defocusing TB to form a
localized soliton. The initial re-phasing generates a DS around
one of the intermediate sites, which then periodically hops
outwards. While the initial site in which the DS forms is a
function of $\Delta\theta$, the hopping period is always the same
for a given $K_{x}$, corresponding to a constant propagation
angle. The good agreement with the experimental data suggests that
the spectral and temporal dynamics are not crucial for the
understanding of the observed interactions.

In conclusion, we have studied experimentally the interactions
among two non-collinear discrete solitons in glass waveguide
arrays. We have demonstrated for the first time that a tilted beam
(TB) launched in a self-defocusing direction, while deviating away
from its normally propagating counterpart (NS), can be
re-localized and routed into any intermediate site by the
nonlinear interaction with the NS. The routing is controlled by
the initial phase difference between the solitons. In simulations,
this behavior was reproduced under several conditions, in which
(i) the nonlinear length is considerably smaller than the
diffraction length; (ii) the NS and TB have comparable input
amplitudes; and (iii) the TB input amplitude is moderate, to
enable soliton steering and to avoid its trapping and instability.
When two on-site excitations exist simultaneously, $\Phi_{n}=A+B$,
the nonlinear Kerr term in the coupled equations (1) acts as a
non-trivial interference term
$|\Phi_{n}|^{2}\Phi_{n}=(|A|^{2}+|B|^{2}+2|A||B|sin\Delta\theta)(A+B)$,
and a new propagation mode that is in balance with the weak
inter-site coupling is obtained. Under the above conditions for
the initial collective excitation, this mode favors localization
of the initially SDF TB, which can indeed be controlled with the
initial phase difference as a single parameter. The result is a
unique recapturing of a defocusing beam into a soliton, which
should have analogs in many other physical systems that have
nontrivial dispersion properties and support localized nonlinear
modes. In addition, the interaction that we observe has potential
applications in all-optical switching devices.

This work was supported by the Israel Science Foundation through
an Excellence-Center grant No. 8006/03 and contract No. 0900017,
and by NSERC in Canada. We thank V. Fleurov and S. Flach for
enlightening discussions. Y.L. appreciates hospitality of
Universit\'{e} du Quebec during his stay.

\end{document}